\documentstyle[twocolumn,prl,aps,floats]{revtex}

\begin{document}
\input{psfig.sty}

\draft
\title{Comment on ``Thermal Effects on the Casimir Force in
the 0.1-5 $\mu$m Range'' by M. Bostr\"om and Bo E. Sernelius
(Physical Review Letters {\bf 84}, pp. 4757-4760 (2000))}

\author{S.K. Lamoreaux}

\address{University of California,
Los Alamos National Laboratory,
Physics Division P-23, M.S. H803, Los Alamos, NM 87545}

\date{\today}

\maketitle

In a recent paper [1] it is shown that simultaneous
consideration of the thermal
and finite conductivity corrections to the Casimir force between
metal plates results in a significant deviation from previous
theoretical results [2] and a recent experiment [3].
In [1], it is argued that the $TE$ electromagnetic mode
does not contribute as $\omega\rightarrow 0$ for a realistic
material with dissipation (see Eq. (6) of [1]).

When the zero-frequency limit is taken in [1], it is asserted that
\begin{equation}
\lim_{\omega\rightarrow 0} (\gamma_1-\gamma_0)=0
\end{equation}
where
$\gamma_{0,1}=q^2+\epsilon_{0,1}(i\omega)\omega^2/c^2$,
0 refers to the space between the plates (vacuum, so
$\epsilon_0(i\omega)=1$)
and 1 refers to the region within the plates.
In [1], the assertion given in Eq. (1) above is based on the idea
that, as $\omega \rightarrow 0$,  $\omega^2\epsilon_1(i\omega)\rightarrow 0$
with the assumed form of the dielectric function, Eq. (6) of [1].

In taking the limit in this manner, some important physics is
missed.  The two evanescent wave numbers, $\gamma_0$ and $\gamma_1$
describe the damping of the electric field in the space between the
plates and within the plates, respectively.  It is well-know that
for static electric fields that the electric field is perpendicular
to a conducting surface (even a poor conductor) and that the field
penetration depth into the conductor is infinitesimally small (e.g., the
field is cancelled due to charge accumulation on the surface).
This implies that $\gamma_1\rightarrow \infty$ when $\omega\rightarrow 0$,
while $\gamma_0$ remains finite.

In taking the zero-frequency limit, the proper result can obtained
in a transparent fashion
if instead we consider $\gamma_1/\gamma_0$ which should diverge at
zero frequency.  The point that was missed in [1] is that
any non-zero $q$ must be associated with a non-zero
$\omega$ because the plates are 
uncharged, and any differential accumulation of
charge on the plate surfaces (e.g., due to thermal fluctuations)
will be dynamic; $q$ and $\omega$ cannot be taken as 
independent integration variables, but are related in
such a way that when $\omega\rightarrow 0$ then $q\rightarrow 0$.
   
Using the substitution $q=\omega\sqrt{p^2-1}/c$ where $p\geq 1$ (as
did Lifshitz [4]; this particular substitution simplifies later
integrals), we find that
\begin{equation}
{\gamma_1\over\gamma_0} ={\sqrt{p^2-1+\epsilon_1(i\omega)}\over p}
\end{equation}
which indeed diverges as $\omega\rightarrow 0$, as expected based on
the simple physical arguments presented earlier, implying that
in this limit
\begin{equation}
G^{TE}=1-e^{-2\gamma_0 d}
\end{equation}  
(which is Eq. (2) of [1]).
Carrying through this substitution
(see [5], pp. 227-228 and pp. 269-271), the previously obtained
result is reproduced [2,4,5] for which there is no significant deviation
between theory and experiment.

\end{document}